\documentclass[aps,prd,twocolumn,groupedaddress]{revtex4}

\usepackage{graphicx}
\usepackage{dcolumn}
\usepackage{bm}

\def\beq{\begin{eqnarray}}
\def\eeq{\end{eqnarray}}
\def\e{{\rm e}}

\newcommand{\ket}[1]{\left| #1 \right\rangle}

\newcommand{\ketss}[3]{\left| #1 \right\rangle_{#2}^{#3}}

\begin{document}


  \title{The clock ambiguity and the emergence of physical laws}


\author{Andreas Albrecht}
\author{Alberto Iglesias}
\affiliation{University of California at Davis\\ Department of
  Physics\\ One Shields Avenue \\ Davis, CA 95616}


\date{\today}

\begin{abstract}
The process of identifying a time variable in time reparameterization
invariant theories results in great ambiguities about the actual laws of
physics described by a given theory.  A theory set up to describe one set of
physical laws can equally well be interpreted as describing any other
laws of physics by making a different choice of time variable or ``clock''.
In this article
we demonstrate how this ``clock ambiguity'' arises and then discuss how one 
might still hope to extract specific predictions about the laws of
physics even when the clock ambiguity is present.  We argue that a
requirement of quasi-separability should   
play a critical role in such an analysis.  As a step in this direction,
we compare the Hamiltonian of a local quantum field theory with a
completely random Hamiltonian.  We find that any random Hamiltonian
(constructed in a sufficiently large space) can yield a ``good
enough'' approximation to a local field theory.  Based on this result we argue
that theories that suffer from the clock
ambiguity may in the end provide a viable fundamental framework for
physics in which locality can be seen as a strongly favored (or
predicted) emergent behavior.  We also speculate on how 
other key aspects of known physics such as gauge symmetries and Poincare
invariance might be predicted to emerge in this framework. 
\end{abstract}

\pacs{}

\maketitle

\section{Introduction}
\label{Sect:I}
In attempts to find a physical description of the universe one has to address 
many issues forced upon us by consistency with quantum mechanics. A 
well-known example is an aspect of time that arises in the
quantization of gravity. In any theory with time reparameterization invariance,
including Einstein gravity, quantization schemes tend to produce
theories in which time is not fundamental, being only  
recovered after some split of the superspace is performed to identify
a time parameter or a ``choice of clock''. In~\cite{Albrecht:1994bg}
it was argued that the freedom to choose a clock leads to profound
ambiguities in the physics that emerges.  In this article we study the 
implications of taking these ambiguities seriously.
Specifically, we consider the fact that the clock ambiguity implies that
completely random choices of unitary evolution of the physical systems
are on an equal physical footing.  A detailed derivation and
discussion of the clock ambiguity is presented in
Section~\ref{Sect:TCA} of this article. 

We examine the possibility that the clock ambiguity is a
fundamental characteristic of physical laws, which forces us to
regard other crucial properties of the physical world such as space,
locality, gravity, gauge symmetries and cosmology as emergent and
approximate. In Section~\ref{Sect:UCFF} we consider how one might
best set up the problem so that the emergence of these properties
could be studied and understood.   

To test the viability of these ideas we compare a random Hamiltonian
with that of a local field theory in Section~\ref{Sect:SFAF}.
Remarkably, we find that in sufficiently large spaces {\em any} random
Hamiltonian appears to give a sufficiently good approximation to a
local field theory to account for the viability of local field theory
as a description of the observed physical world.  Note
that our starting point is an arbitrary random {\em Hamiltonian} (not
an arbitrary Hamiltonian {\em density}). We make 
no initial assumption about the existence of space, locality, etc. We
are claiming that these properties can quite generally be seen as
emergent from a random Hamiltonian. 

A priori, placing all possible Hamiltonians on an equal footing seem to
be hopelessly in conflict with standard approaches
to physics. Certainly one possible outcome of this work is to cause
the abandonment of at least one of the assumptions that go into stating the
clock ambiguity.  We discuss this possibility in Section
\ref{Sect:TCA:D}, but note that this outcome would be 
significant, in that our assumptions are ones that are widely used in
quantum cosmology.

Taking the clock ambiguity at face value, it would seem that
extracting the known physical laws from a situation where all
possibilities for those laws are initially given equal weight would involve
eliminating most of those possibilities for one (probably anthropic) reason or 
another.  Thus we feel our result from Section~\ref{Sect:SFAF} is
extremely interesting.  It shows that good approximations to local
field theories can be found very generically in randomly chosen
Hamiltonians. We take this as an indication that a framework for
fundamental physics with the clock ambiguity rooted firmly in its
foundations may not be nearly as problematic as it first seems. 
We feel our work offers a strong motivation for taking
such a framework seriously and making further efforts to explore its
ultimate viability.

\section{The clock ambiguity}
\label{Sect:TCA}
\subsection{Statement of the clock ambiguity}
\label{Sect:TCA:SOTCA}
Here we review the clock ambiguity as discussed in
\cite{Albrecht:1994bg}. Our starting point is a standard approach to
time in quantum gravity whereby time is defined internally~\footnote{
The approach we use here is discussed very nicely in section
6.2 of Isham's comprehensive review~\cite{Isham:1992ms} of time in quantum
gravity, where it is referred to as ``internal time using conditional
probabilities''.  We accept all the positions Isham identifies as
required for a consistent application of this approach (including a post-
Everett~\cite{Everett:1957hd} approach to quantum measurement).  We
also note that our willingness to assume a superspace that is discrete
and finite (an additional assumption not used by Isham) allows us to
sidestep  normalizability issues pointed out 
by Isham. This approach to time is sometimes called ``Page Wooters
Time'' \cite{PageWooters}}.
Any time reparameterization invariant theory (including
General Relativity)  
has the property that the Hamiltonian is zero~\cite{Arnowitt:1960es} and the 
reparameterization can be viewed as a gauge transformation generated by the 
first-class constraint $H$. In a covariant approach to quantization this 
feature is imposed as a constraint equation 
\begin{equation}
  H \ket{\psi} = 0~,
\end{equation}
on physical states $\ket{\psi}$ in a ``superspace'' that includes both matter
and metric degrees of freedom.  Time evolution is regained by
identifying some degree of freedom (or ``subsystem'') as the ``clock''
and evaluating correlations between the rest of the universe and the
state of the clock subsystem.  For
example, several classic papers on quantum
cosmology~\cite{Lapchinsky:1979fd,Hartle:1983ai,Banks:1984cw,Banks:1984np,Halliwell:1984eu,Fischler:1985ky}
use the cosmic scale factor $a$ as the clock degree of freedom. 

This approach gives a practical way forward when considering time in
quantum gravity, but it also make intuitive sense. The process of
identifying a subsystem of the world as a ``clock'' and noting the
passage of time in terms of correlations with the clock subsystem
gives a good operational picture of how we actually work with time
in realistic situations. 

In order to describe the clock ambiguity we assume the superspace may be taken
to be discrete and finite.  Although continuous quantities (fields,
the metric, and spacetime itself) are usually used to described
physics, observations do not rule out the idea that these continuous
quantities are just approximations to a system that is fundamentally
discrete and finite. (This fact has been used by others seeking a
discrete and finite fundamental description of physics, see for
example~\cite{Banks:2001ge,Banks:2003ta,Banks:2004eb,Banks:2004cw,Banks:2004vg,Buniy:2005iw,Dyson:2002pf}.) 
Assuming discreteness and finiteness will make our mathematical
manipulations simple.  

Formally, if $S$ designates the superspace, then the identification of
a clock subsystem involves designating the clock ($C$) and ``rest''
($R$) subspaces of $S$ so that
\begin{equation}
S =  C\otimes R~.
\label{CRpartition}
\end{equation}  
Let 
\begin{equation}
\left\{ \ketss{t_i}{C}{ }\right\}
\label{clockbasis}
\end{equation}
be a basis which spans the clock space (eigenstates of the time
operator in the language of Isham~\cite{Isham:1992ms}) and let basis
\begin{equation}
\left\{\ketss{j}{R}{}\right\}
\end{equation}
span $R$.   The tensor product of the of the bases spanning $C$ and
$R$ spans $S$, so any state $\ketss{\psi}{S}{}$ in superspace can be written
\begin{equation}
\ketss{\psi}{S}{} =
\sum_{ij}\alpha_{ij}\ketss{t_i}{C}{}\ketss{j}{R}{}
\label{CRexpand}
\end{equation}
where $\alpha_{ij}$ are the expansion coefficients.   One can the sum
up the $j$'s for a fixed time (fixed $i$) to get the states
\begin{equation}
\ketss{\phi_i}{R}{} \equiv \sum_j\alpha_{ij}\ketss{j}{R}{}.
\label{phidef}
\end{equation}
One can then rewrite Eqn.~\ref{CRexpand} as
\begin{equation}
\ketss{\psi}{S}{} = \sum_i \ketss{t_i}{C}{}\ketss{\phi_i}{R}{}.
\label{tphiexpand}
\end{equation}
The state $\ketss{\psi(t_i)}{R}{}$ of subsystem $R$ at time $t_i$ is
determined by conditioning
(projecting) on clock state $\ketss{t_i}{C}{}$, giving~\footnote{
Any concern one might have about the need to respect the 
normalization condition ${ }_S\!\langle \psi(t_i) |\psi(t_i) \rangle_S
$ can be resolved by carefully formulating questions in terms of
conditional probabilities}
\begin{equation}
\ketss{\psi(t_i)}{R}{} = \ketss{\phi_i}{R}{}.
\label{psioft}
\end{equation}
So far we have just summarized a standard approach to time in quantum
gravity using a formal discrete notation that will be useful in what
follows. 

Now we present the argument from~\cite{Albrecht:1994bg} that suitably changing
the choice of clock subsystem can lead to a description of an
arbitrary system experiencing arbitrary time evolution. 

To start with, we note that all the information about the state and
the time evolution is contained in the $\alpha_{ij}$'s of
Eqn.~\ref{CRexpand}.   We will show that by choosing different clock
subsystems we can get arbitrary $\alpha_{ij}$'s, which will then
correspond to arbitrary states undergoing arbitrary time evolution. 

It will be helpful to relabel the tensor product basis for $S$
used in Eqn.~\ref{CRexpand} with a single index.  This involves
defining some mapping $k(i,j)$ that uniquely assigns an index $k$ to
each pair $(i,j)$ so one can write
\begin{equation}
  \ketss{k(i,j)}{S}{} \equiv \ketss{i(k)}{C}{}\ketss{j(k)}{R}{}
\label{kSdef}
\end{equation}
Then one can write 
\begin{equation}
\ketss{\psi}{S}{} = \sum_k \alpha_k \ketss{k}{S}{}.
\label{kSexpand}
\end{equation}
where 
\begin{equation}
  \alpha_k\equiv \alpha_{i(k),j(k)}
\label{akdef}
\end{equation}
where the functions $i(k)$ and $j(k)$ simply invert the mapping
$k(i,j)$. In this notation, arbitrary $\alpha_{ij}$'s corresponds to
arbitrary $\alpha_k$'s. 

But arbitrary $\alpha_k$'s are easy to attain through a change of
basis. To see this, suppose one starts with a particular vector given
by Eqn.~\ref{CRexpand} or equivalently Eqn.~\ref{kSexpand}, and would
like to demonstrate an alternative choice of clock describing a
specific different state and time evolution.  The goal is to construct
a new set of subsystems
\begin{equation}
  S = C'\otimes R'
\label{newCR}
\end{equation}
and the appropriate bases in $C'$ and $R'$ so that 
\begin{equation}
\ketss{\psi}{S}{} =
\sum_{ij}\beta_{ij}\ketss{t_i}{C}{\prime}\ketss{j}{R}{\prime}
\label{newCRexpand}
\end{equation}
where the $\beta_{ij}$'s give the required information about the state
and its time evolution (just as $\alpha_{ij}$ did for the original
case).  The first step is to use the same function $k(i,j)$ discussed
above to construct 
\begin{equation}
  \beta_k \equiv \beta_{i(k),j(k)}
\end{equation}
and then consider a new vector in the superspace
\begin{equation}
  \ketss{\psi'}{S}{} \equiv \beta_k \ketss{k}{S}{}
\label{psiprime}
\end{equation}
(note that here the original superspace basis $\left\{\ketss{k}{S}{}\right\}$
is used). Now consider a unitary~\footnote{We choose ${\bf M}$ to be 
unitary in order to preserve basis normalizations when we operate (later) 
on basis vectors with it.  The overall norm of $\ketss{\psi}{S}{}$ has no 
significance.} transformation ${\bf M}$ that
transforms $\ketss{\psi}{S}{}$ into $\ketss{\psi'}{S}{}$
\begin{equation}
  {\bf{M}}\ketss{\psi}{S}{} =   \ketss{\psi'}{S}{} 
\label{Mdef}
\end{equation}
(it should be always possible to find at least one such
transformation). Operating on both sides of Eqn.~\ref{Mdef} with
$\bf{M}^{-1}$ gives
\begin{equation}
  \ketss{\psi}{S}{} =  {\bf{M}}^{-1} \ketss{\psi'}{S}{}  = \sum_k
  \beta_k {\bf{M}}^{-1}\ketss{k}{S}{}.
\end{equation}
If one then defines a new basis
\begin{equation}
  \ketss{k}{S}{\prime} \equiv  {\bf{M}}^{-1}\ketss{k}{S}{}
\end{equation}
one gets
\begin{equation}
  \ketss{\psi}{S}{} =  \sum_k \beta_k \ketss{k}{S}{\prime}
\label{betak}
\end{equation}
The desired $C'$ and $R'$ subsystems are
constructed using the inverse of the mapping function $k(i,j)$ (the
same one used above) to give
\begin{equation}
  \ketss{i(k),j(k)}{S}{\prime} \equiv  \ketss{i}{C^\prime}{}
  \ketss{j}{R^\prime}{} =  \ketss{k}{S}{\prime}
\label{CRprimeresult}
\end{equation}
leading to 
\begin{equation}
  \ketss{\psi}{S}{} =  \sum_{ij} \beta_{ij} \ketss{i}{C^\prime}{} \ketss{j}{R^\prime}{}
\label{betaij}
\end{equation}
which is the desired result. 

Basically we have used the fact that while a different state evolving
under a different Hamiltonian would seem to correspond to a different
state $\ketss{\psi'}{S}{}$ in superspace, it could just as well be
seen as the {\em same} state in superspace expressed in a different
basis (corresponding to a different subdivision of the system into
``Clock'' and ``Rest'' subspaces).  We have used $\ketss{\psi'}{S}{}$ 
as well as the mapping
function $k(i,j)$ to explicitly demonstrate how such a new basis can
be constructed for $S$.   

The implication of our result is that given
that all possible clocks corresponding to  all possible time
evolutions  can be demonstrated to exist, a physicist trying to
interpret $\ketss{\psi}{S}{}$ from scratch is equally likely to try
any one of these clock subsystems, thus placing all possible types of
evolution on an equal footing.
Specifically, a single state in
superspace can be interpreted as any initial state evolving under any Hamiltonian. 

\subsection{Discussion}
\label{Sect:TCA:D}
The result in Section~\ref{Sect:TCA:SOTCA} is radical, but it seems
to be an inevitable consequence of standard ideas about quantum
gravity.  One could take the standard model of particle physics (or
one's favorite extension thereof), combine it with gravity and construct the
corresponding superspace. Then someone else could come along using
the exact same rules of interpretation you use, but by merely choosing a
different clock could come up instead with a world described by the
old $O(3)$ model of weak interactions, the MSSM, technicolor, or
something wildly different from any of these.

It is important to emphasize that the clock ambiguity is {\em not}
equivalent to the statement that it is possible to choose terrible
clock subsystems (for example, a firefly) by whose measure the evolution of
the universe appears highly irregular (although the clock ambiguity does incidentally include these
cases).  The most important implication of the clock ambiguity is that
it also includes a multitude of arbitrarily good clocks which describe
the universe evolving under very well defined and ``sensible'' physical
laws. The clock ambiguity tells us that there is nothing about the form of
superspace nor the state which we choose in superspace to give a 
preference of one set of physical laws over another, no mater how hard
we may try at the outset to build such a preference into the
formalism.

One possible response to our analysis is that one or another of our
assumptions is wrong.  For example, perhaps there is something truly
precious about the continua we use to construct theories of
fundamental physics, and our discrete and finite treatment misses some
key point. One could also choose to reject the superspace
formalism outright as is done for example by Banks {\em et al.} in
\cite{Banks:2002wr}.  

One might also object that we should not be allowed freedom to choose
a clock subsystem arbitrarily but should stubbornly stick to the one
originally designated. That objection seems to run up against
commonly held views in quantum cosmology.  For example in ``eternal
inflation''~\cite{Linde:1986fd} the system in some regimes is completely 
dominated by
quantum fluctuations. Which combination of some ``fundamental'' states
and operators in superspace end up representing actual semiclassical
observables (including time) will depend strongly on which piece of
the wavefunction one is looking at (or which quantum fluctuations one
is following). The idea that one must dig through a more formally
constructed space to select observables based on their actual behavior
is widely used.  See for example \cite{Page:2006ys} and also
\cite{Rovelli:1990ph,GMHsfi89}. 
Indeed, in our analysis in the previous
section it is not just the clock but all observables that are changed
when going from one picture to another.

One might even question the use of the covariant approach. The alternative 
being the fixing of the original reparameterization symmetry at the 
classical level by imposing a 
gauge condition, {\it i.e.}, a choice of reference time. While such a choice
of an external time is suitable for the study of subsystems, with negligible 
interaction with the environment (that would naturally set such a reference)
it presents no advantage when dealing with the universe as a whole. In fact, 
in this case, the absence of external reference and subsequent
arbitrariness in  
the choice of gauge, we believe to be analogous to the arbitrariness in the 
choice of clock subsystem of Section~\ref{Sect:TCA:SOTCA}. An early
discussion of this point of view can be found in
\cite{Arnowitt:1959xx}.  However, these 
authors chose a framework that was too restrictive to expose the full
clock ambiguity.  

The original paper on the clock ambiguity~\cite{Albrecht:1994bg} gives
further discussion of the objections that might be raised about our
formalism, and gives responses to these.  Also, Isham's 
review~\cite{Isham:1992ms} (especially section 6.2) gives a good account of
some pros and cons of this formalism (although Isham does not use our
assumption of discreteness and finiteness).

Our main position on all of this is that the clock ambiguity is a very
important topic.  If careful consideration of the
clock ambiguity leads to rejection of some of the starting
assumptions, we feel that would be a significant outcome since these
assumptions are currently widely accepted, especially by those who
work with quantum cosmology. 

The rest of this paper focuses on another possible outcome, namely
that the formalism used above really does describe fundamental
physics.  In that case, the clock ambiguity is something we need to
face head on.  We investigate possible ways forward
under that assumption.

\section{Useful conditions for finding a good clock}
\label{Sect:UCFF}

\subsection{Overview}
\label{Sect:UCFF:O}

The clock ambiguity seems to leave us very little to work with.  Can a
fundamental physical theory that appears to put all possible states evolving
under all possible Hamiltonians on an equal footing make any concrete
statements about the nature of physical laws?   If the clock ambiguity
is a real feature of fundamental physics, then the fact that the world
is so understandable in terms of specific physical laws must mean that
there really must be a way forward. 

In this section we consider some possible ways preferences for
specific physical laws could emerge in this picture.  We continue the
approach developed in~\cite{Albrecht:1994bg}, where fundamental aspects of our experience as
observers are identified and considered as selection criteria in
choosing our Hamiltonian and state from among all possible ones.  

This approach seems to fall under the broad category of 
``anthropic reasoning''(as used for example in
\cite{BTanthropic,Weinberg1987,Tegmark:2005dy,Tegmark:1997in}) 
but as emphasized in~\cite{Albrecht:1994bg} and developed further 
in~\cite{Albrecht:2002uz}, our approach should be seen as a natural application
of the conditional probability analysis that underlies most
applications of theoretical physics to actual observations.  Probably
the most controversial aspect of anthropic reasoning (and rightly so
in our view) involves  attempts to incorporate general ``physical conditions
necessary for life'' as conditions in conditional probability
statements.  We do not believe that we (or any 
other physicists) really know the general physical conditions
necessary for life so we refrain from speculating on those
here. Instead, we consider what appear to be general features of our
interaction with the rest of the universe.  These features are just as
essential to inanimate observers (such as automated data acquisition
systems) as they are to us.

If this picture is to succeed, we expect to eventually
reach the point where more familiar conditions are applied (such as
observations of the electron mass fixing its value in quantum
electrodynamics). However the current picture is so far removed from
that stage that we only emphasize here more exotic conditions that
could offer a glimmer of hope that some sort of preference for
specific physical laws could emerge. 

\subsection{Time independence of the Hamiltonian}
\label{Sect:UCFF:TIOT}
A striking aspect of the clock ambiguity is that it
gives no a-priori preference for evolution under a time {\em independent}
Hamiltonian.  In the notation of Section~\ref{Sect:TCA}, the
Hamiltonian should generate steps between adjacent discrete times.
Specifically, 
\begin{equation}
  \ketss{\psi(t_{i+1})}{R}{} = -i\hbar(t_{i+1}-t_i){\bf
  H(t_i)}\ketss{\psi(t_i)}{R}{}.
\label{Hdef}
\end{equation}
But since   $\ketss{\psi(t_{i+1})}{R}{}$ and $\ketss{\psi(t_i)}{R}{}$
are just defined separately by the $\alpha_{i+1,j}$'s and
$\alpha_{i,j}$'s respectively (see Eqn.~\ref{phidef} and
Eqn.~\ref{psioft}), 
and since the clock ambiguity allows
one to consider on an equal footing all possible $\alpha_{i+1,j}$'s,
regardless of the values of the $\alpha_{ij}$'s, there is no a-priori
reason to assume any particular relationship between ${\bf H}(t_i)$
and ${\bf H}(t_{i+1})$.

Certainly the constancy of the laws of physics over time appears to be
a critical part of our experience as observers.  We count on such
constancy to learn about our environments (both with our minds, and
through genetic evolution) and reap the benefits from the knowledge we
gain. So there seems to be hope that this aspect of our existence as
observers could be related to (approximate) time independence of
${\bf H}$, but at this point do not have a quantitative analysis to
offer.

\subsection{Hermiticity of the Hamiltonian}

In contrast to the time dependence of the Hamiltonian, the
self-adjoint property of ${\bf H}$ is realized in a 
straightforward way.  In standard quantum mechanics the Hamiltonian is
taken to be Hermitian in order to ensure that time evolution is
unitary.  Unitary evolution allows wavefunctions normalized to unit
total probability to remain so normalized as time evolves.  As
emphasized by Isham~\cite{Isham:1992ms}, the formalism we use here
makes explicit use of conditional probabilities.  For example, you could 
calculated the probability of measuring a particle at position $x$ {\em
  given} that the clock is in state $\ketss{t_i}{C}{}$.  To do this
you would project onto the $\ketss{t_i}{C}{}$ state and normalize the
answer so that the total probability assigned to all possible outcomes
of the measurement (given the time projection) is unity. The overall
normalization of the wavefunction before projection is unimportant
when formulating conditional probabilities. 

One way to put this is that given that only conditional probability
questions will be posed, all possible time evolutions will be treated
in a way that makes them effectively unitary.  Any non-unitary aspect of
the $\alpha_{ij}$'s will drop out of the final analysis.

\subsection{Quasi-separability and locality}
\label{Sect:UCFF:QSAL}

\subsubsection{Overview}
\label{Sect:UCFF:QSAL:O}

Our experience in the universe is characterized by the fact
that we are minuscule subsystems of the universe that are able to survive and
even thrive with respect to our interactions with the rest of the
universe.  We are able to keep our interactions with the rest of the
universe from destroying us for a period of time (with luck, several
decades).  Furthermore, the state of the rest of the universe can be
accounted for in a highly simplified manner that allows us to model the rest
of the universe with our tiny little brains in a way that usually is
sufficient for our survival: Even though a bus has many more
microscopic physical degrees of freedom than our brains, on average we
manage to model the behavior of those degrees of freedom sufficiently
well to avoid being hurt by the bus, and even to utilize it for
transportation. Planets, stars and galaxies (with vastly more internal
degrees of freedom than a bus) are even easier to handle. 

This behavior is extremely different from that which these various
subsystems (galaxies, stars, planets, buses and us) would experience
under evolution given by an arbitrary Hamiltonian. In the most general
case one would not expect the interactions between these subsystems to
be particularly weak or predictable. In fact, there would be no reason
to expect the interactions to be at all subdominant to these
subsystems' self-interactions.  The generic result would seem to be
subsystems that are rapidly torn apart by their interactions with the
rest of the universe in a manner that prevents them from keeping any
identity as subsystems. 

In this section we note that the observed peaceful coexistence of subsystems
reflects the quasi-separable nature of the Hamiltonian that actually
governs our world. We take the point of view that the
quasi-separability is sufficiently important (and sufficiently
non-generic) that it should be taken as a key condition to impose as we
search among all arbitrary Hamiltonians for ones that might be
relevant to the physical world we experience. 

We then note that the quasi-separability we actually experience is
closely related to the {\em  locality} that is manifested by the fundamental
physical laws we observe. We then speculate on the degree to which imposing the
quasi-separable requirement on arbitrary Hamiltonians could strongly
favor local physics, perhaps even sufficiently strongly to favor
Hamiltonians approximating local quantum field theories with local
gauge symmetries and gravity. 

\subsubsection{Locality}
\label{Sect:UCFF:QSAL:L}

In our experience, the key to the quasi-separable nature of our world
is the locality of physics. As long as we occupy different locations
from the buses, planets, stars and galaxies we have a reasonable shot
at not being destroyed by them.  Formally, this locality comes about because
Hamiltonians that describe known physics take the form 
\begin{equation}
  H = \int {\cal H}(x) d^3x.
\label{localH}
\end{equation}
This is certainly very far from the most general case.  A general
Hamiltonian would allow arbitrary interactions between matter at any two
points. Even within the local formalism, there are two long-range
forces: gravity and electromagnetism. The overall neutrality of the
universe cuts back greatly on the impact of electromagnetism, and the
overall (homogeneous and isotropic) state of the universe limits the
impact of the long-range gravitational forces between objects.  Also,
the timescale for gravity to have its full impact (such as earth's
orbit decaying and plunging us into the sun) is long compared to
timescales that interest us. 

The critical role locality plays in realizing the quasi-separability
that is so important to us leads us to speculate that locality could turn out to
be a ``generic'' way for quasi-separability to emerge as one sifts
through arbitrary Hamiltonians.  It could be that when
quasi-separability is sought that optimizes the evolution of small
successful observers it tends to naturally lend itself to interpretation
in a ``local'' language.  Since locality is a crucial piece of the
construction of quantum field theory, perhaps one could even use
arguments such as these ``derive'' quantum field theory as a foundation
of our understanding of matter. 

A key part of locality is the definition of space and of distances
between points in that space.  A more general realization of these
features will come 
about if one allows distances in the space to be defined in terms of
an arbitrary metric $g_{ij}$. When one sifts through
random systems and selects out ones that exhibit locality, presumably
many more examples will turn up with complicated metrics than with
simple ones.  Such a tendency toward non-trivial spatial
metrics might lay the groundwork for Einstein
gravity to emerge in this picture. 

\subsubsection{The speed of light}
\label{Sect:UCFF:QSAL:TSOL}

Another key aspect of known physics is the bounding of all speeds by
the finite speed of light. In the picture we describe here, if we
enforce locality then propagation speeds should be finite (but not
necessarily equal) in all directions. At each point, and in each
direction, there will be a maximum propagation speed experienced by
certain degrees of freedom. Perhaps it will turn out to be natural to
define all other propagation speeds relative to this maximum speed.  

The more quantitative analysis of section \ref{Sect:SFAF} suggests 
an interesting perspective on the emergence of full Poincare
invariance, which we discuss briefly in section \ref{Sect:SFAF:D}.

\subsection{Spin, Statistics and Gauge Symmetry}
 
If Poincare invariance and locality do indeed emerge in this picture,
presumably this will lead to the emergence of field operators in
various representation of the Lorentz group. Since the spin statistics
relation is understood to be a consequence of
locality~\cite{Pauli:1939xp,Pauli:1940}, the critical role of locality in
our picture should enforce the usual spin statistics relation. 

Since one can argue that gauge symmetries are necessary for the
consistency of massless spin one fields (see for 
example~\cite{Weinberg:1980kq} and Chapter 8 of~\cite{WeinbergFT}), the 
random appearance of some massless spin one
fields might be all it takes for gauge symmetries to emerge in this
picture~\footnote{
Despite clear differences, we note there seems to be an interesting
overlap between these ideas about 
gauge symmetry (and also about the emergence of general relativity
mentioned above) and ideas
about emergence that are developing in the context of string
theory~\cite{Seiberg:2006wf}}. 
One possible outcome is that the probabilities for the emergence of
particles in different representations of the Lorentz group lead to
preferences of particular gauge symmetries over others. 

\subsection{The arrow of time and the state of the universe}
\label{Sect:UCFF:TAOTATSOTU}

The statistical foundations of the thermodynamic arrow of time make it
natural to associate the arrow of time with special (low entropy) initial
conditions of the universe. This point was first made in a
modern cosmological context by Penrose~\cite{Penrose:1980ge}.  Starting with
Boltzmann~\cite{Boltzmann}, some physicists have been interested in a
cosmological picture with an eternal equilibrium state that
occasionally fluctuates so 
as to produce a region of (temporarily) low entropy.  The regions of
increasing entropy associated with these fluctuations then become
candidates to describe our world~\cite{Albrecht:1994bg,Page:1983uh,
Dyson:2002pf,Albrecht:2002uz,Albrecht:2004ke}.
In general this picture is believed to suffer from the ``Boltzmann Brain''
problem~\cite{Boltzmann,Rees:1997,Albrecht:2004ke}, whereby small fluctuations 
containing only one observer for a brief moment dominate the
predictions. 
One of us has  
argued in~\cite{Albrecht:2002uz} and~\cite{Albrecht:2004ke} that the 
Boltzmann Brain problem could be
resolved by a period of cosmic inflation, and that the resolution of
the Boltzmann Brain problem is in fact one of the key attractive
features offered by
inflation.   

We find it interesting to compare the picture developed in this paper
with Boltzmann's ``fluctuation from equilibrium'' picture.  Formally,
one might think that since we consider a finite system there should
be quasi-periodic recurrences of the sort that were considered 
in~\cite{Dyson:2002pf} and~\cite{Albrecht:2004ke}.  However, just because
a subsystem $C$ has ``good clock'' behavior for a sufficiently long
period to describe our observations does not mean it would be a good
clock over a {\em complete} set clock states $\ketss{t_i}{C}{}$ that
span $C$.  It is quite possible that most realistic depictions of our  
universe in this formalism would involve the breakdown of the ``good
clock'' behavior at some point outside of the observed
domain~\cite{Salecker:1957be,Isham:1992ms,Hartle:1988kt,Unruh:1989db,Banks:2002wr}.  
That would make it hard to define Boltzmann's fluctuating equilibrium
state over ``eternity'' (a.k.a. a complete 
recurrence time). This is an intriguing point,
especially since in the Boltzmann picture it seems a bit of a waste to
have time well defined over the extremely long equilibrium period when
it is of no real use to us without a thermodynamic arrow. 

Still, it is quite possible that something similar to the arguments 
in~\cite{Albrecht:2004ke} will apply in formalism described in this
paper.  In that case the need for a thermodynamic arrow of time will
not only play a key role as a condition for searching for realistic
clocks, it will also play a critical part in biasing the initial
conditions of the observed universe toward those that were subject to
an early period of inflation.

\subsection{Dimensionality of space, classicality and other considerations}
\label{Sect:UCFF:DOSC}
There are a number of other factors that could have an important
impact on the selection of a good clock. For example the term ``classicality'' is
often applied to the various combinations of phenomena (including the
dominance of quantum path integrals by saddle points and the
stability of a measurement apparatus after a quantum
measurement).  More generally, one needs spacetime itself to behave in
a classical manner to accurately describe the world we see around
us. Many of the phenomena associated with classicality have
already been mentioned in this section (for example the emergence of space,
locality and the thermodynamic arrow of time).  It is possible that requiring
additional aspects of physics that lead to classical behavior produces
additional constraints in the sort of analysis envisioned
here.

Also, in a picture where space is emergent, one naturally wonders if
there is any preferences for one number of space dimensions over
another.  Several ideas along these lines have been been put forward
over the years (see for example~\cite{Tegmark:2002uz}).  We will return 
briefly to this issue in
section~\ref{Sect:SFAF:D}, but so far we do not yet feel we have a
compelling argument that a particular number of space dimensions would
be favored.  

\section{Searching for a field theory in a random Hamiltonian}
\label{Sect:SFAF}

\subsection{Overview}
\label{Sect:SFAF:O}

In Section~\ref{Sect:UCFF} we considered possible ways forward to extract
meaningful physics out of quantum gravity, despite the clock ambiguity
described in Section~\ref{Sect:TCA}.  There is clearly far to go if that
approach is to really bear fruit. In this section we take a 
``reverse engineered'' approach and ask to what extent the known laws 
of physics might match on to a random Hamiltonian. 

The critical point of comparison is the eigenvalue spectrum:  We
draw a Hamiltonian at random by choosing a random
clock. If its eigenvalue spectrum matches one corresponding to that
of the standard model of particle physics then we are ``done'' in the
sense that there is nothing in principle stopping us from carefully
identifying the requisite field operators, observables, etc. that describe
the theory in the usual way in terms of the eigenstates of $H$.  We
don't know how easy this would be in practice, but we do not believe
it would run into any issues of principle. 

Having stated our approach, we discuss (in the next section) some
general results from the theory of random Hamiltonians. Then, in
Section~\ref{Sect:SFAF:DOSO} we consider the eigenvalue spectrum
of a free field theory (as a first step toward the eigenvalue spectrum
of a full interacting theory).
In Section~\ref{Sect:SFAF:A} we attempt a comparison between the field
theory spectrum and that of a random Hamiltonian.  Although at first
glance comparison seems futile, we suggest an intriguing way forward 
which appears to hold considerable promise. 

\subsection{Properties of random Hamiltonians}
\label{Sect:SFAF:PORH}

\begin{figure}
\includegraphics[width=3.5in]{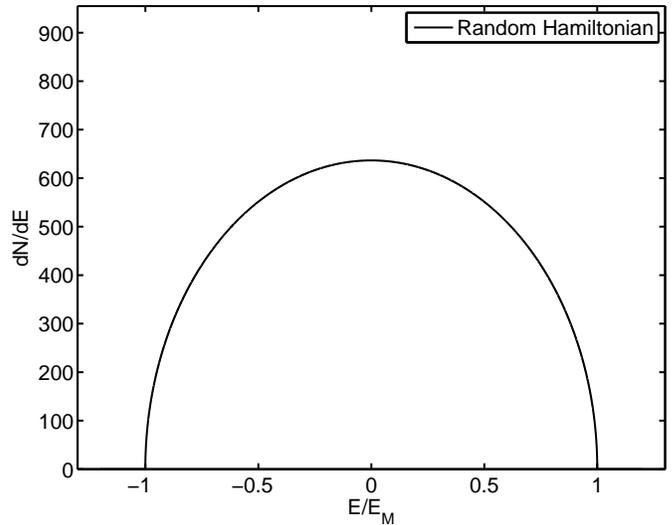}
\caption{\label{fig:dndeR} A plot of the density of states for a
  random matrix as given by the Wigner semicircle rule. (We plot
  Eqn. \ref{eqn:Wigner} with $E_M = 1$ and $N_H = 1000$)}
\end{figure}

There is an extensive literature on random Hamiltonians (see
\cite{Mehta}, \cite{Brezin:1993qg} and references therein).  The basic
idea is to select each matrix element of the 
Hamiltonian from some distribution and look at the ensemble that
emerges. It turns out that a wide variety of such random Hamiltonians
end up obeying the ``Wigner semicircle rule''
\begin{equation}
{dN\over dE}=\left\{\begin{array}{ll}{2N_H\over\pi E_M} 
\sqrt{1-\left({E\over E_M}\right)^2}~~~& |E|<E_M~,\\ 0 &{\rm otherwise.}
\end{array}\right.
\label{eqn:Wigner}
\end{equation}
(We derive this in Appendix~\ref{apa} for the Gaussian case.) Here
$E_M$ is the maximum eigenvalue and 
$N_H$ is the size of the random Hamiltonian. One can wonder if 
there might be subtleties in the
process of generating random Hamiltonians through the ``choice of clock'' 
process that do not generate eigenvalue spectra of exactly this
form.  Thus we consider a slightly generalized form
\begin{equation}
{dN\over dE}=\left\{\begin{array}{ll} a {N_H\over E_m} 
\left(1-\left({E\over 
E_M}\right)^\beta\right)^\gamma~~~&|E|<E_M~,\\ 
0 &{\rm otherwise.}\end{array}\right.
\label{randNgen}
\end{equation}
so we can see how possible variations on the standard form might
affect our final results.   An
illustrative example from this class of functions is depicted in
Fig.~\ref{fig:dndeR}. 

\subsection{Density of states of a free field theory}
\label{Sect:SFAF:DOSO}
\begin{figure}
\includegraphics[width=3.5in]{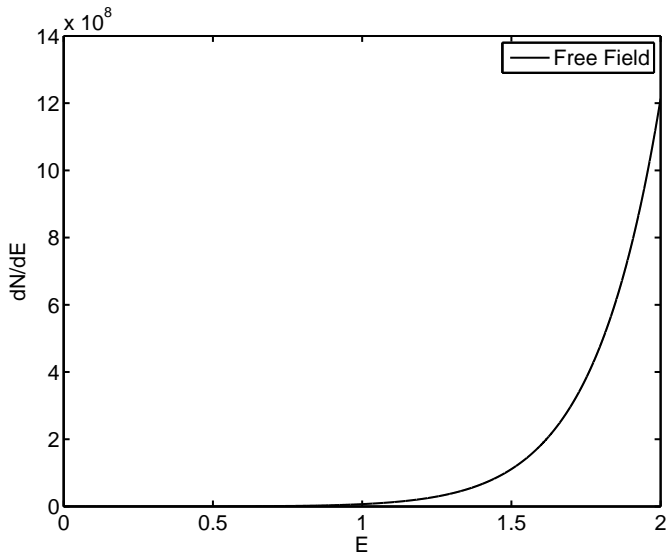}
\caption{\label{fig:dndeF}
A plot of the field theory density of states given by
Eqn.~\ref{dNFFT} (using $B= b = 1$,
$\alpha = 1/2$ and $\Delta k = 0.001$). }
\end{figure}

Ideally, at this point we would write down a formula for $dN/dE$ for
the standard model of particle physics and compare it with
Eqn.~\ref{randNgen}.  Since we do not know this function, we seek some
initial insights by considering $dN/dE$ for a free field theory.  We
are not aware of prior calculations of this quantity either, but in
$1+1$ dimensions we show in Appendix~\ref{apb} that 
\begin{equation}\label{Nft}
{dN\over dE}
\sim{1\over4\sqrt{3}E}\exp\left\{\pi\sqrt{2E\over 3\Delta k}\right\}~,
 ~~~~E\gg \Delta k~,
\end{equation}
for a free boson. The quantity $\Delta k$ reflects the fact that we
have regulated the field theory by putting it in a box of size $ L =
2\pi/\Delta k$.  A similar expression is
also found for free Fermions.  
Thus, we consider in this case the following generalization,
\begin{equation}\label{dNFFT}
{dN\over dE}=
{B\over E}\exp\left\{b\left({E\over\Delta k}\right)^\alpha\right\}~,
\end{equation}
for large $E$, which for $b=1/2$ contains, as special cases, the $1+1$
expressions of Appendix~\ref{apb} and the higher dimensional 
generalization proposed by Verlinde
in~\cite{Verlinde:2000wg}\footnote{If one considers instead an
  extensive form for the entropy one finds a different extension of
  this formula to higher dimensions (see for example
  \cite{Banks:1999az}). We conisder this case in \cite{Albrecht:2008xx}} . When
both Fermi and Bose fields are combined, the density of states is
dominated by the Bose fields which is why Eqn. \ref{dNFFT}
reflects the Bose form.  An 
illustrative example of this type of function is shown in Fig.~\ref{fig:dndeF}.

\subsection{Analysis}
\label{Sect:SFAF:A}
The forms of Eqn.~\ref{randNgen} and Eqn.~\ref{dNFFT} (also depicted
in Figs.~\ref{fig:dndeR} and~\ref{fig:dndeF}) are dramatically
different.  At first look this suggests that finding a field theory by
randomly choosing a clock (and thus generating a random Hamiltonian)
is a very unrewarding endeavor.  At best, only Hamiltonians on highly
exponentially suppressed tails of the random distribution might give
the needed eigenvalue spectrum.  It is perhaps not
surprising that the ``conditions for a good clock'' discussed in
section \ref{Sect:UCFF} would seek out atypical cases. Still, the striking
difference between these two functional forms seems to indicate
how extremely selective these conditions would have to be in order to
allow this whole approach to succeed.  

However, there might be a much easier way forward from here. The key
is to consider the fact that we only actually explore $dN/dE$ of the
universe in some relatively narrow range of energies $\Delta E$ around
a mean energy $E_0$ (the ``energy of the Universe'').  Given this, 
we can ask if  Eqn.~\ref{randNgen} and Eqn.~\ref{dNFFT} can look
similar in that particular range of energies.

We start with Eqn.~\ref{randNgen}, the generalized form of the Wigner
semicircle result 
\begin{equation}
{dN\over dE}=\left\{\begin{array}{ll} a {N\over E_m} 
\left(1-\left({E\over 
E_M}\right)^\beta\right)^\gamma~~~&|E|<E_M~,\\ 
0 &{\rm otherwise.}\end{array}\right.
\end{equation}
and the generalized form of the free field theory result
(Eqn.~\ref{dNFFT})
\beq
\frac{dN_F}{dE}=\frac{B}{E}
\exp\left\{b\left(\frac{E}{\Delta k}\right)^\alpha\right\}~. 
\label{eqn:FTgen}
\eeq
We attempt to equate these two equations order by order in a Taylor
expansion around $E_0$.

First we note that we are not trying to consider gravity
at this point. As discussed in Section~\ref{Sect:UCFF:QSAL:L} gravity could
potentially emerge through deeper insights into the emergence of
locality (and thus a metric).  Without gravity, we can assume in this
very simpleminded comparison that the overall zero point of $E$ does
not have any physical meaning (it just causes an unobservable overall
phase shift in the time evolution).  Thus we allow a zero point shift
$E¬_S$ when comparing the Wigner and free field theory expressions.
Specifically we relate $E_R$, the energy in the Wigner random Hamiltonian
expression (Eqn.~\ref{randNgen}) to $E_F$, the energy in the free field
theory expression (Eqn.~\ref{dNFFT}) according to  
\beq
E_R  = E_F  - E_S ~.
\eeq
We keep Eqn. \ref{eqn:FTgen} unchanged and absorb the shift into
Eqn. \ref{randNgen} to give
\begin{equation}
{dN\over dE}=\left\{\begin{array}{ll} a {N\over E_m} 
\left(1-\left({{E-E_S}\over 
E_M}\right)^\beta\right)^\gamma~~~&|E|<E_M~,\\ 
0 &{\rm otherwise.}\end{array}\right.
\label{randNgenS}
\end{equation}

We then Taylor expand each of the expressions for the density of
states around the central energy $E_0$. Expanding the generalized Wigner
formula (Eqn.~\ref{randNgenS}) gives
\begin{widetext}
\beq
   \frac{dN_R}{dE}&=&
   a\frac{N_H }{E_M}\left(1-\left(\frac{E_0-E_S}{E_M}\right)^\beta\right)^\gamma   
   \left\{1-\beta\gamma Q
\frac{\Delta E}{E_0}
+\frac{1}{2}\gamma Q
\left[\left(\gamma-1\right) Q
-\left(\beta-1\right)\right]
\left(\frac{\Delta E}{E_0} \right)^2+\ldots\right\} ~,
\eeq
where $1/Q\equiv ((E_0-E_S)/E_M)^\beta-1$. In turn, the field theory formula 
(Eqn. \ref{dNFFT}) gives
\beq
   \frac{dN_F}{dE}&=&
   B\frac{1}{E_0 }\exp \left\{ b\left( \frac{E_0 }{\Delta k}
   \right)^\alpha \right\}\left[ 1 + \alpha b\left(
   \frac{E_0}{\Delta k}\right)^\alpha\frac{\Delta E}{E_0}
+\frac{1}{2}\alpha b
\left(\frac{E_0}{\Delta k}\right)^\alpha
   \left[\alpha b\left(\frac{E_0}{\Delta k}\right)^\alpha
   + \left(\alpha-1\right)\right]\left(\frac{\Delta E}{E_0}\right)^2
   +  \ldots \right] 
\eeq
\end{widetext}

Demanding equality at 0th
order and solving the resulting expression for $N_H$ leads to  
\beq
N_H=\frac{B}{a}\frac{E_M }{E_0 }
\left[1-\left( \frac{E_0-E_S}{E_M}\right)^\beta\right]^{-\gamma}
\!\!\exp\left[b\left(\frac{E_0}{\Delta k}\right)^\alpha\right]
\label{eqn:Zeq}
\eeq
This expression is completely dominated by the
exponential (even though we will soon argue that the quantity in
square brackets is extremely small).  Thus 0th order equality of the
two densities of states just sets the size of the space of the random
Hamiltonian to be some specific exponentially large number.  It seems
reasonable to regard Eqn.~\ref{eqn:Zeq} as a fundamental relation in
our formalism. We note that since data only give an upper bound on the
field theory regulator $\Delta k$, Eqn.~\ref{eqn:Zeq} should really be
seen as giving a lower bound on $N_H$.

Requiring equality between the free field and the generalized Winger expression
at first order (as well as at zeroth order) leads to
\beq
 - \beta \gamma \frac{{E_0 }}{{E_0  - E_S }}
\frac{{\left( {\frac{{E_0  - E_S }}{{E_m }}} \right)^\beta }}
{{\left\{ {1 - \left[ {\frac{{E_0  - E_S }}{{E_M }}} \right]^\beta }
    \right\}}} 
= \alpha b\left( {\frac{{E_0 }}{{\Delta k}}} \right)^\alpha~.  
\label{eqn:Feq}
\eeq
The right hand side is expected to be an exponentially large quantity
(the ratio of the energy of the universe field theory regulator
$\Delta k$).
To achieve equality for Eqn. \ref{eqn:Feq} requires the quantity in square
brackets to be exponentially close to unity.   This leads to  
\beq
E_S  = E_0  - E_M \left( {1 - \varepsilon } \right)~.
\label{eqn:ESvalue}
\eeq
Here $\varepsilon$ (equal to the quantity in curly brackets in
Eqn. \ref{eqn:Feq}) must be exponentially small (determined implicitly
from Eqn.~\ref{eqn:Feq}) and must be positive (by the nature of the Wigner
formula).  Also, one must have  
\beq
E_S>E_0~,
\eeq
in order to get the overall sign right.  It seems reasonable to also take
Eqn.~\ref{eqn:Feq} as a fundamental relation for our scheme.

Since we are considering a finite system, there will be a finite gap $\Delta_G$
between energy eigenvalues which can be estimated by 
\beq
\begin{array}{*{20}c}
   {\Delta _G } &  =  & {\left( {\left. {\frac{{dN_R }}{{dE}}}
   \right|_{E_0 } } \right)^{ - 1}  = \left( {\left. {\frac{{dN_F
   }}{{dE}}} \right|_{E_0 } } \right)^{ - 1} }  \\ 
   {} &  =  & {\frac{{E_0 }}{B}\exp \left\{ { - b\left( {c\frac{{E_0
   }}{{\Delta k}}} \right)^\alpha  } \right\}}~.  \\ 
\end{array}
\label{DeltaGDef}
\eeq
Comparing this gap with the field theory regulator gives 
\beq
\frac{{\Delta _G }}{{\Delta k}} 
= \frac{{E_0 }}{{\Delta k}}\frac{1}{B}\exp 
\left\{ { - b\left( {\frac{{E_0 }}{{\Delta k}}} \right)^\alpha  }
\right\}~.
\label{DeltaGdk}
\eeq
This suggests that as long as we set
$\Delta k$ small enough to respect the phenomenological successes of
continuum field 
theory, any effects due to the finiteness of the random Hamiltonian
are exponentially suppressed and are unlikely to be unobservable.
Similar arguments suggest that random fluctuations in density of
states due to specific realizations of the random Hamiltonian will be
highly subdominant, although we have not exhaustively investigated
this question.  

It is also irresistible to note that Eqn.~\ref{eqn:ESvalue} includes
$E_0$, the energy of the
universe, in the energy offset. In the absence of a specific notion of how gravity emerges
in this picture it is really too early to speculate, but we can't help
but wonder if this offset might end up offering an interesting insight
into the cosmological constant. 

Equality of the two densities of states at second order gives 
\beq\label{eqn:Seq}
{1 \over \gamma} \alpha b\left( {\frac{{E_0 }}{{\Delta k}}}
\right)^\alpha   = \alpha  - 1 + \beta~,
\eeq
where we have also used the conditions of equality at zeroth and first
order (Eqns.~\ref{eqn:Zeq} and~\ref{eqn:Feq}).    The left hand side of
Eqn.~\ref{eqn:Seq} is generally an extremely large number (the ratio of
the energy of the universe to the $k$-space regulator of the
field theory).  
The right hand side is definitely of order unity.  Clearly one
cannot expect to impose exact equality at second order.
Specifically, the second order difference between the two
densities of states is given by
\beq
\left( \frac{dN_F}{dE}  -
  \frac{dN_R}{dE} \right)_2 
 & =&\frac{1}{2}\frac{B}{{E_0 }}\exp \left\{ {b\left( {\frac{{E_0
  }}{{\Delta k}}} \right)^\alpha  } \right\}\alpha b\left( {\frac{{E_0
  }}{{\Delta k}}} \right)^\alpha  \nonumber \\
&&\hskip-1cm\times\left( {\alpha  - 1 -
  \frac{1}{\gamma }\alpha b\left( {\frac{{E_0 }}{{\Delta k}}}
  \right)^\alpha   + \beta } \right)\left( {\frac{{\Delta E}}{{E_0 }}}
  \right)^2  \nonumber\\  
 & \approx & \left. {\frac{dN}{dE}} \right|_{E_0 } \left( {\alpha
  b\left( \frac{E_0 }{\Delta k} \right)^\alpha  } \right)^2
  \left( \frac{\Delta E}{E_0 } \right)^2 \nonumber \\  
  &\approx& \left. \frac{dN}{dE} \right|_{E_0 } \left( {\left(
  \frac{E_0 }{\Delta k} \right)^\alpha  } \right)^2 \left(
  \frac{\Delta E}{E_0 } \right)^2  
\eeq
where we have dropped subdominant terms as
well as factors of order unity to reach the final line.  This leads to
possible fractional corrections to 
the density of states given by
\beq
\frac{\Delta \frac{dN}{dE}}{\left. \frac{dN}{dE}
    \right|_{E_0}} \approx \left( \left( \frac{E_0 }{\Delta
	k} \right)^\alpha  \frac{\Delta
      E}{E_0} \right)^2  +{\cal O}\left( \left( \Delta E \right)^3 
\right)~, 
\eeq
and we label the second order piece as
\beq
\Delta_2  \equiv \left( {\left( {\frac{{E_0 }}{{\Delta k}}}
  \right)^\alpha  \frac{{\Delta E}}{{E_0 }}} \right)^2 ~.
\label{eqn:Delta2}
\eeq

Now we consider the overall size of $\Delta_2$.  We take $E_0$ to be
the energy of the observed universe, namely 
\beq
  {E_0 } &  =  & {\rho _c R_H^3  \approx \left( {H_0^2
   m_P^2 } \right)H_0^{ - 3}  = m_P^2 H_0^{ - 1} } \nonumber  \\ 
    &  \approx  & {\frac{{10^{19} GeV}}{{10^{ - 42}
   GeV}}m_P  = 10^{61} m_P  = 10^{80} GeV} ~,
\label{eqn:E_0}
\eeq
where $R_H$ is the Hubble length, $H_0$ is the Hubble constant today,
and $m_P\approx 10^{19} GeV$ is the Planck mass.   We
have chosen the zero point of the energy as the point with zero particle
excitations, so $E_0$ in Eqn. \ref{eqn:E_0} is the correct value to use
in the field theory density of states.  Taking care of the
contribution of the dark energy to this estimate could lead to $O(1)$
correction factors that do not concern us here. 

The quantity $\Delta k$ gives the scale of discreteness for the field
theory.  The fact that so far we have no evidence for discreteness
suggests some pretty low upper bounds on $\Delta k$. We consider one
value of $\Delta k$ given by the current bound on the photon mass 
$\Delta k=m_\gamma \approx 10^{-25}GeV $~\cite{PDG,Ryutov:1997}.  We also consider
$\Delta k = H_0 \approx 10^{-42}GeV$, the wavenumber of a wave the
size of the observed universe. 

The quantity $\Delta E$ should give the range of energy eigenvalues 
over which we expect field theory to a give good representation of
physics. That is, the range over which we hope the two densities of
states will coincide to a good approximation. Physically, a lower
bound on $\Delta E$ is set by the shortest time $\delta t$ over which we
successfully model observed phenomena using field theory. The two are
related by 
\begin{equation}
  \Delta E \geq {\hbar \over \delta t}
\end{equation}
We can look to ultra high energy cosmic rays 
($\delta t^{-1} \approx 10^{11}GeV$)
or the highest energy elementary particle accelerators 
($\delta t^{-1}\approx 10^3 GeV$) to set values of $\delta
t$~\footnote{There may be subtleties regarding the extent to which these
  values of $\delta t$ reflect genuine field theory effects vs. the
  behavior of some collective coordinate.  Those subtleties should be examined
  carefully before attaching significance to an exact value of $\delta
t$.  However, this issue does not appear to be important for the very
broad points being discussed in this paper.} 

Table~\ref{Delta2} gives the values of $\Delta_2$ which correspond to
a selection of different values for the quantities in
Eqn.~\ref{eqn:Delta2} mentioned in the above discussion.  The upshot
is that as long as $\alpha=1/2$ 
(the value given in our formula for $1+1$ and also in
Verlinde's suggested generalization to higher dimensions) the second
order fractional correction to the density of states is very small
($\Delta_2 \approx 10^{-8}$ or even $\Delta_2 \approx 10^{-24.5}$).
This small deviation between the free field theory and Wigner formulas
might account for interactions seen within the context of field theory
or perhaps yet-to-be-observed deviations from field theoretical
behavior in the real physical world. Also, the field theory density of
states assumes a Minkowski space, so deviations of the real spacetime
from the Minkowski form will show up as a discrepancies in this
analysis. (We do not yet have a calculation of any of these
phenomena.)  Also, we note that the small values of $\Delta_2$
encourage us to believe that contributions form the higher order terms
in $\Delta E$ will be even smaller for the values of $\Delta E$ that
are of interest.  

\begin{table}
\begin{ruledtabular}
\begin{tabular}{cccc}
$\alpha$ & $\Delta k$ &$ \Delta E$ & $\Delta_2$ \\ \hline
$1/2$  & $m_\gamma = 10^{-25}GeV $  & $1 TeV$ & $10^{-24.5}$ \\
$1/2$  & $m_\gamma = 10^{-25}GeV $  & $10^{11} GeV$ & $10^{-16.5}$ \\
$1/2$  & $H_0 = 10^{-42}GeV $  & $1 TeV$ & $10^{-16}$ \\
$1/2$  & $H_0 = 10^{-42}GeV $  & $10^{11} GeV$ & $10^{-8}$ \\
$1$  & $m_\gamma = 10^{-25}GeV $  & $1 TeV$ & $10^{28}$ \\
$1$  & $m_\gamma = 10^{-25}GeV $  & $10^{11} GeV$ & $10^{36}$ \\
$1$  & $H_0 = 10^{-42}GeV $  & $1 TeV$ & $10^{45}$ \\
$1$  & $H_0 = 10^{-42}GeV $  & $10^{11} GeV$ & $10^{53}$ \\
$2$  & $m_\gamma = 10^{-25}GeV $  & $1 TeV$ & $10^{133}$ \\
$2$  & $m_\gamma = 10^{-25}GeV $  & $10^{11} GeV$ & $10^{141}$ \\
$2$  & $H_0 = 10^{-42}GeV $  & $1 TeV$ & $10^{167}$ \\
$2$  & $H_0 = 10^{-42}GeV $  & $10^{11} GeV$ & $10^{175}$ \\
\end{tabular}
\end{ruledtabular}
\caption{\label{Delta2}Value of $\Delta_2$ for different choices of the 
exponent $\alpha$ in Eqn.~\ref{eqn:FTgen}, observable energy range $\Delta E$ 
and field theory regulator $\Delta k$. As long as $\alpha = 1/2$, $\Delta_2$
takes on small values, suggesting the random Hamiltonian is giving a
good approximation to the field theory and also validating the Taylor
expansion.} 
\end{table}

We have thus shown that it is possible to find 
``field theory like behavior'' in the density of states of any random
Hamiltonian. The key is to only look for equivalence between the field
theory and the random Hamiltonian over a {\em finite range} of
energies eigenvalues $\Delta E$.  Specifically, we have show that for sufficiently large spaces
($N_H$ set by Eqn. \ref{eqn:Zeq}) the offset energy ($E_S$) can be
suitably chosen (using Eqn. \ref{eqn:Feq}) so that the field theory
and random densities of states are identical to zeroth and first orders
and only differ by a small amount (given by $\Delta_2$ from
Eqn. \ref{eqn:Delta2}) at second order.   

\begin{figure}
\includegraphics[width=3.5in]{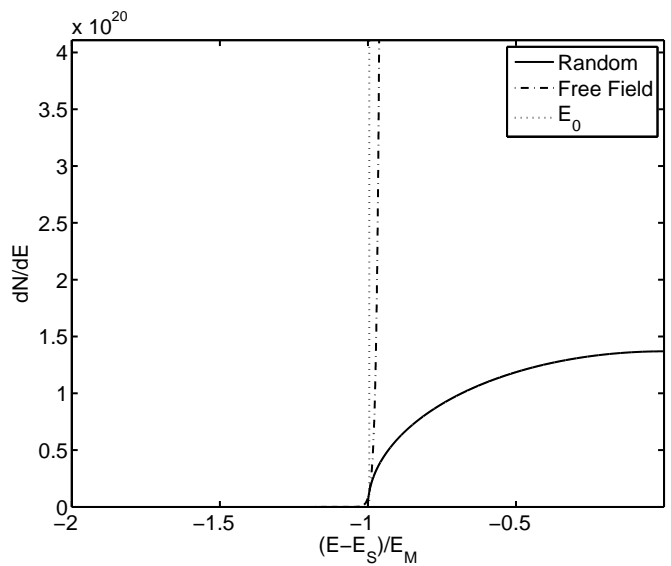}
\caption{\label{fig:both} This figure plots curves for
  $dN_R/dE$   $dN_F/dE$ from Eqns.~\ref{randNgenS}
  and~\ref{dNFFT} respectively. Equations~\ref{eqn:Zeq} and~\ref{eqn:Feq} have
  been imposed to cause the zeroth and first order terms in Taylor
  expansions to be equal at $E_0$ (marked by the vertical line).  The
  point of coincidence is chosen to be close to the edge of the
  circle, as discussed around Eqn. \ref{eqn:ESvalue}, but for easier viewing the
  value of $\varepsilon$ (which measures the proximity to the circle edge) in this plot is
  much larger than the exponential small values discussed in the
  text.}
\end{figure}

\begin{figure}
\includegraphics[width=3.5in]{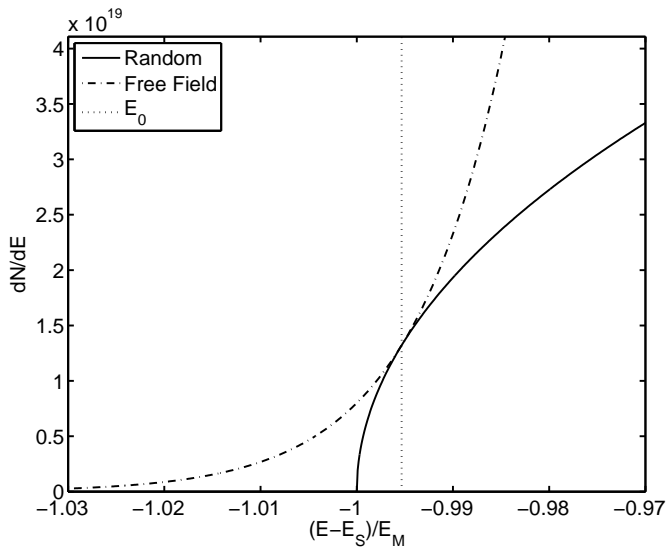}
\caption{\label{fig:bothzoom}The same plot as Fig.~\ref{fig:both} but
  zoomed in to show more detail where
  the curves coincide}
\end{figure}

Figures \ref{fig:both} and \ref{fig:bothzoom} give an illustration
of how the two densities of states can be made to coincide over a
finite range of energies, despite their radically different overall
shapes depicted in Figs. \ref{fig:dndeR} and \ref{fig:dndeF}. For this
illustrative 
example we have chosen $a=2/\pi$, $E_M = 10$,   $B=1$, $b=1$,
$\alpha=1/2$ and $\Delta k = 0.001$. For these parameters solving
Eqn. \ref{eqn:Zeq} gives $N_H = 2.15\times 10^{21}$ and solving
Eqn. \ref{eqn:Feq} implicitly gives $E_S = 11.9533$ which we use 
for these plots. 

\subsection{Discussion}
\label{Sect:SFAF:D}
Our results suggest the following picture: A physicist analyzes a
state in superspace by choosing a clock subsystem at random,
resulting in the particular story about the time evolution of the rest
of the space which is associated with that clock.   The
physicist then poses the question ``how well can this time evolution
accommodate tiny observer-subsystems such as ourselves'' .  The idea here
is that the clock choices that maximize the viability of
observer-subsystems give laws of physics more likely to be the ones we
observe (assuming we are ``randomly selected'' observer-subsystems
found in our superspace). 

By the arguments of section \ref{Sect:UCFF:TIOT} clocks which give
Hamiltonians which are (at least 
approximately) time independent should be favored.
Since separability is very helpful to observer-subsystems, the laws of
physics which maximize separability, namely local field theory, are
the ones that are most favored by this selection process.  
The results from the previous subsection tell us that any random
Hamiltonian can be interpreted as a local field to a very good
approximation.  Thus the search for separability appears 
to {\em predict} the emergence of local field theory. 

One then must go about analyzing the implications of the deviations
from a free field theory represented by $\Delta_2$ and terms higher
order in $\Delta E$.   We speculate that the goal of optimizing a local
interpretation (and thus good separability) will be very significant in
interpreting these deviations from free field theory. We expect it to
lead to allocations of these deviations among a variety of corrections
to the simple ``free field in Minkowski space'' starting point.  These
corrections include deviations from Minkowski space (non-trivial
metrics), the allocation of degrees of freedom among different
particles with the spin statistics relation imposed 
(as usual) by the need for locality, as well as interactions between
the different particles.  As argued in section~\ref{Sect:UCFF:TIOT},
this analysis also might be expected to reveal emergent gauge
symmetries.  We also note that while the approach of section~\ref{Sect:SFAF:A}
was to set the free field and random densities of states precisely equal at
zeroth and first order, optimizing the overall interpretation of
the random matrix as an interacting field theory with gravity might
involve slightly relaxing strict equality at those orders. 

A very interesting issue is the degree to which the process of assigning local
interpretations to the $\Delta_2$ etc. results in any predictive
power. Perhaps this process could eventually be understood to give some
very powerful predictions about particle content, symmetries
etc. based on the same sort of statistical arguments that
operate to give arbitrary random Hamiltonians a unique spectrum
given by the Wigner semicircle\footnote{We are taking a statistical
  approach to the laws of physics. We note that similar ideas have come up
  recently in the context of the string theory
  landscape\cite{Kachru:2003aw} (see for example 
  \cite{Kofman:2004yc}). Here we are drawing our laws of physics
  from a much more general starting point than the string theory 
  landscape (but not as general as Tegmark's ``mathematical
  democracy''\cite{Tegmark:2002uz}).  We also note that that a
  concrete realization of a relation between random matrix 
  models in the large $N$ limit and certain four dimensional
  supersymmetric gauge theories can be found in the work of Dijkgraaf
  and Vafa \cite{Dijkgraaf:2002dh} and further generalizations.}.     

One intriguing direction is to consider the role of Poincare
invariance in this picture.  In our analysis, Poincare invariance
has an impact on form of the density of states of a field theory due
to its role in 
defining the dispersion relation of free particles.  It is possible
that local theories that have dispersion relations inconsistent with
Poincare invariance will not exhibit the good behavior under the
Taylor expansion noted in this paper.  For example they may manifest
the exponentially large second order ``corrections'' seen in some cases
discussed above (see for example the $\alpha =1$ and $\alpha = 2$ cases in
Table \ref{Delta2}).  It is conceivable that such considerations could
lead to a sharp preference for Poincare invariant physics, a possibility we
are currently actively investigating. 

When setting up this analysis, we introduced extra parameters to
produce generalized forms of the density of states for free field
theory and a random Hamiltonian (these parameters are $a$, $\gamma$,
$\beta$, $B$, $b$ and $\alpha$). 
These parameters were introduced to evaluate the robustness of our
analysis.  At the broad level of our current discussion which is focused on
the degree to which random Hamiltonians can approximate a local field
theory, the only one of these parameters that seems to matter is
$\alpha$.  The other parameters show up in the equations, but changing
their values does not change our main points.  

We expect the number of
space dimensions to enter through the parameter $b$ in the field theory
density of states (Eqn. \ref{eqn:FTgen}).   The lack of sensitivity to $b$
in our current discussion means we have yet uncover any factors that
would prefer one number of space dimensions over another.  As
discussed in section~\ref{Sect:UCFF:DOSC}, other considerations might lead to
such a preference. 

\section{Summary and Conclusions}
\label{Sect:C}

We are used to doing physics by stating the physical laws which we
believe may be true, and then calculating predictions based on those laws in
order to test them against observations of the physical world. The
clock ambiguity appears to completely undermine this approach to
physics.  In time reparameterization invariant theories such as
Einstein gravity, the process of identifying a time variable
creates the clock ambiguity.  Even if one carefully sets up the
system to reflect a particular set of physical laws, the exact same
state in superspace can be viewed from the point of view of a
different time variable or clock which causes the system to exhibit
completely different laws of physics. 

A reasonable response to this observation is to reject one or more of the
assumptions that go into demonstrating the clock ambiguity.  To this
end we have carefully identified these assumptions in this paper, and
we have argued that rejecting any of these assumptions would be an
interesting development since the assumptions we used are widely
accepted among physicists. 

Most of this article has focused on the possibility that the clock
ambiguity is a central feature of fundamental
physics, a feature that we are going to have to learn to live with.  
We first considered the type of analysis that might allow some 
concrete predictions about the physical world to emerge, despite the
profound ambiguities introduced by the choice of clock.  Specifically,
we envisioned an approach where the fact that we are tiny subsystems
of the entire 
universe which are able survive and thrive plays a key role in
selecting the type of physical laws we observe.
We argued that
laws of physics that allow subsystems to do well will preferentially
be those that are observed and analyzed by such subsystems.  

We identified a number of features of physical laws that would
promote the success of small subsystems, and gave special attention to
quasi-separability of the Hamiltonian.  This is the feature that
allows small subsystems to interact much more strongly with themselves
than with their environment and thus keep their identities.
Quasi-separability also allows subsystems to successfully model their 
environment based on a simple set of collective coordinates without
knowing much about every single degree of freedom.   We noted that in our
experience, it is the locality of the laws of physics that leads to
the quasi-separability on which we so heavily depend and we argued that 
local physics (as expressed by local field theories) seems
to be the optimal way of achieving maximal amounts of
quasi-separability.  

In order to probe this line of thinking in a more quantitative manner,
we investigated the extent to which a local quantum field theory could
be approximated by a completely random Hamiltonian. Remarkably, we
discovered that if the random Hamiltonian is constructed in a large
enough space it can {\em always} approximate a free local field theory
to a sufficient degree.  Here ``sufficient'' means part of the
spectrum of eigenvalues (or density of states) can coincide with that
of a free field theory over a range of eigenvalues. The range of
eigenvalues, centered on $E_0$, the energy of the observed universe,
need only be sufficient to reproduce all observed phenomena that we
believe are explained by local field theories. This is very different
from matching the full eigenvalue 
spectrum of a field theory.  For example, the field theory ground
state on which so much of the formal construction of field theory is
based is not part of the spectrum at all.  None the less, such an
approximation to a true field theory may be sufficient to account for
the success of field theoretic models of the physical world.

Initially our picture seemed to be one in which one would reject an
enormous fraction of all possible clock subsystems based on the
inappropriateness of the corresponding Hamiltonian evolution.  We
thought that surely
conditions such as quasi-separability must be very far from 
universal.   Our result that good approximations to field theory
can always be found in sufficiently large random Hamiltonians changes
this story considerably.  It now appears that any random Hamiltonian
can be optimized for quasi-separability by constructing a local field
theoretic interpretation.  In this sense local field theories might
actually be seen as a prediction. 

Many questions still remain.  
How well is the
possible time dependence of the Hamiltonian constrained in our
picture? 
Is there some further optimization
process that can lead to concrete predictions about gauge symmetries,
Poincare invariance, general relativity, etc.?  
We have speculated along these lines but so far we do not
have concrete results.   Still, we find it intriguing that 
local field theories are much easier to come by than we initially
expected. We feel this result offers hope that a framework for fundamental
physics which suffers from the clock ambiguity may in the end prove
viable.  

One of us has argued elsewhere~\cite{Albrecht:2004ke}
that statistical arguments offer a much more powerful approach to
cosmological initial 
conditions than the more traditional approach of making an ad hoc
statement of preference.  It is possible that eventually the
statistical approach to laws of physics described in this paper could
achieve that sort of standing.  While the outcome is still far from 
clear, we feel the results in this paper motivate a further
investigation of this possibility.   

\begin{acknowledgments}
We are grateful to Cliff Burgess, Steve Carlip, Nemanja Kaloper and
Constantinos Skordis for a number of helpful technical discussions.
We thank Jesus De Loera and Andrew Waldron for discussions of the
field theory density of states (and we specifically thank Waldron for
sharing with us his alternative derivation of the Bose results from
Appendix B). Pedro Ferreira gave helpful feedback on a talk by one of
us\cite{Albrecht:EverettAt50} which helped sharpen the discussion in
this paper.  Damien Martin and Tony Zee pointed out the Wigner
semicircle result to us.  We also thank Tom Banks, David Deutsch,
Willy Fischler, Manoj Kaplinghat, Lloyd Knox, Don Page, Max Tegmark
and David Wallace for more general discussions which helped to
stimulate this work.  This work was supported in part by DOE Grant
DE-FG03-91ER40674  
\end{acknowledgments}

\appendix
\section{Derivation of the Wigner semicircle distribution}\label{apa}

We consider an ensemble of random hermitian
$N_H\times N_H$ matrices representing the possible Hamiltonians that
describe the evolution of states arising in different choices of
clock. As we argue in Section \ref{Sect:SFAF}, when comparing a random
Hamiltonian with the laws of physics as we know them the key point of
comparison is the eigenvalue spectrum.  

\begin{figure}
\includegraphics[width=3.5in]{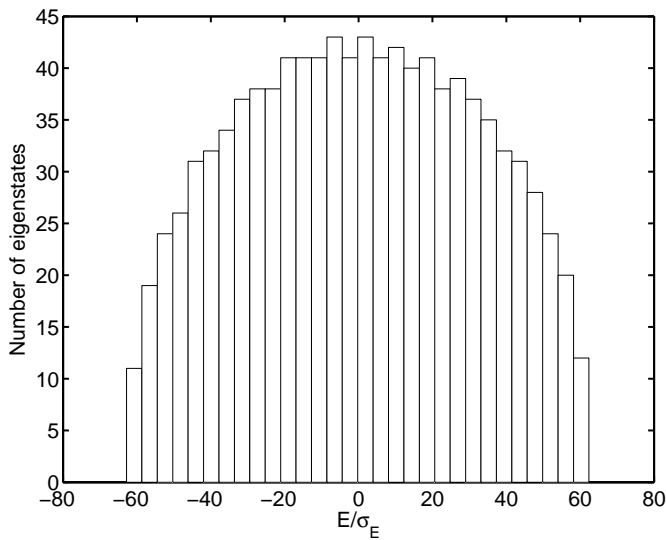}
\caption{\label{fig:Wigner} A histogram of the eigenvalues of a
  $1000\times1000$ Hermitian matrix where the real and imaginary parts
  of each matrix element are drawn from a normal distribution of zero mean
  and width $\sigma_E$ }
\end{figure}

In the limit of large matrices, the eigenvalue spectrum (or density
of states) approaches a unique form.  As an illustration,
Fig.~\ref{fig:Wigner} shows a histogram of the eigenvalues of a
$1000\times 1000$ Hermitian matrix where the real and imaginary parts
of each matrix element were drawn from a normal distribution of width
$\sigma_E$. 
As $N_H$ becomes larger, the fluctuations settle down and the
eigenvalue spectrum (or density of states $dN/dE$) approaches the form
given by the ``Wigner semicircle rule'' (Eqn.~\ref{eqn:Wigner}). 

A derivation of this result (see, for example,~\cite{Mehta}) starts by
considering a random Gaussian distribution of the matrix elements of H
with flat measure $d^{N^2_H}\!H$ over the $N^2_H$ real variables ${\rm Re}
H_{ij}$, ${\rm Im} H_{ij}$ to form the partition function,
\begin{equation}\label{pf1} 
Z=\int d^{N^2_H}\!H~ {\rm e}^{-{1\over2\sigma^2}{\rm tr}~H^2}~.
\end{equation}
The partition function is then conveniently rewritten in terms of
independent variables (the $N_H$ eigenvalues $E_i$ of $H$ and the
elements of the  matrix $U$ that brings $H$ to diagonal form $\Lambda$)
by inserting $1=\int dU \delta(UHU^\dagger-\Lambda)\Delta^2(E)$ where
$\Delta(E)$ is the Fadeev-Popov determinant of Vandermonde form
$\prod_{i<j}(E_j-E_i)$. Integrating over $H$ first and then factoring
out the $U$ integration, \begin{equation}\label{pf2} Z=\int\prod_i
  dE_i~\Delta^2(E)~{\rm e}^{-{1\over 2\sigma^2}\sum_iE^2_i}~, \end{equation}
which has a stationary point at $E_i=2\sigma^2\sum^\prime_j(E_i-E_j)^{-1}$. In
the large $N_H$ limit, a continuous distribution of eigenvalues can be
taken instead
\begin{equation}
E_i=\sqrt{N_H} E(i/N_H)~,
\end{equation}
spread over the interval ($-E_M,E_M$) with variance $\sigma^2=E_M^2/N$ and 
density $dN/dE$ such that
\begin{equation}
{1 \over N_H}\int_{-E_M}^{E_M}dE {dN\over dE}=1~.
\end{equation}
The stationary condition is now
\begin{equation} {1\over 2} E=\int_{-E_M}^{E_M} dE'
{dN\over dE'}{E_M^2\over 2N_H} {1\over E-E'}~,
\end{equation} solved by the Wigner semicircle distribution:
\begin{equation}
{dN\over dE}=\left\{\begin{array}{ll}{2N_H\over\pi E^2_M}
\sqrt{E_M^2-E^2}~~~&
|E|<E_M
~,\\ 0 &{\rm otherwise.}\end{array}\right.
\end{equation}

\section{Density of states for a free field theory}\label{apb}

Here we discuss an example in which the density of states can be 
computed analytically: a massless free field theory in $1+1$ dimensions with 
coordinates $(\sigma,~t)$.
In this case the degeneracy of states at a given energy level can be obtained 
by studying the appropriate generating function.

\subsection{Bosons}

{}Let us first consider a boson confined to an interval of the spacial
dimension,
$\sigma \in [0,\pi]$. The resulting free theory has a discreet spectrum,
with
mode decomposition of the form,
\beq
\phi=\sum_{n\in Z}\!{}^\prime~{1\over\sqrt{n}}a_n\e^{-i n t}\sin n\sigma~,
\eeq
and states with arbitrary occupation number $N_n$ for each mode,
$|N_1,N_2,\cdots\rangle$, with energy $E=\Delta k\sum_{n=1}n~N_n$ (the
energy eigenvalue spacing $\Delta k$ equals one in the present case).

The generating function for this system is given by
\beq\label{Zb}
Z_B={\rm tr}~\e^{\beta E}=\sum_{E=1}^\infty d_E ~ z^E~,
\eeq
where we have defined $z=\e^\beta$, $d_E$ is the degeneracy of states of
energy E (the quantity we are
interested in) and the trace is taken over the state space.

On the one hand, an exact expression is available for $Z_B$,
\beq
Z_B=\prod_{n=1}\sum_{N_n=0}\left(z^n\right)^{N_n}=\prod_{n=1}{1\over 1-z^n}=
z^{-{1\over 24}}\eta(\tau)~,
\eeq
where $\eta$ stands for the Dedekind function and $\tau=\beta/2\pi i$. On
the
other, $d_E$ is easily expressible, from Eqn. \ref{Zb}, as a contour
integral,
\beq\label{con}
d_E={1\over 2\pi i}\oint {Z_B(z)\over z^{E+1}}dz~.
\eeq
By noting that the integrand in Eqn.~\ref{con} is sharply peaked around $z=1$
the
integral can be estimated by a saddle point approximation. The value of
$Z(z\to 1)$ can be deduced from the Hardy-Ramanujan formula that exploits
the
modular property of the Dedekind function
($\eta(-1/\tau)=(-i\tau)^{1/2}\eta(\tau)$):
\beq\label{HR}
{1\over Z_B(z)}=
\sqrt{-2\pi\over\log z}z^{-{1\over 24}}q^{1\over 12}{1\over Z_B(q^2)}~,~~~
q=\e^{2\pi^2\over \log z}~,
\eeq
and realizing that $z\to 1$ corresponds to $q\to 0$ and that $Z_B(0)=1$.
Therefore, the large $E$ asymptotic behavior of $d_E$ is found by
considering
the following approximation to Eqn. \ref{con},
\beq
d_E\sim{1\over 2\pi i}\oint {1\over z^{E+1}}\sqrt{-\log z}~
\e^{-{\pi^2\over 6\log z}}
dz~.
\eeq
The integrand has a stationary point at $\log z=-\pi/{\sqrt{6(E+1)}}$
($z\sim 1$ for large $E$) that readily gives the asymptotic value of
$d_E$:
\beq
d_E \equiv {dN \over dE} \sim{1\over 4\sqrt{3}E}\e ^{\sqrt{2E\over 3\Delta k}\pi}~, ~~~~E\gg
\Delta k~,
\label{dEBose}
\eeq
where we have reinserted, to facilitate the generalization of this
formula, the
spacing between energy eigenvalues $\Delta
k$ (equal to 1 in this case due to the interval of $\sigma$ chosen).

In the case of a compactified spacial dimension, {\it i.e.}, imposing
periodicity at the boundaries, the left and right-moving modes become
independent and the generating function is, therefore, a product of the
two
factors ($Z=|Z_B|^2$).

\subsection{Fermions}

{} Let us consider now a free fermion $\psi$ in $1+1$ dimensions with 
right(left)-moving components $\psi_{-(+)}$,
\beq
\psi=\left(\begin{array}{c} \psi_- \\ \psi_+\end{array}\right)~,
\eeq
and choose, for example, periodic boundary conditions
$\psi_+=\psi_-$ at $\sigma=0$ and $\pi$.
The mode expansions are 
\beq
\psi_\pm=\sum_{n\in Z}\!{}^\prime~ b_n \e^{-i n (t\pm\sigma)}~,
\eeq
and the space of states is labeled by the occupation numbers of each mode as 
in the bosonic case $|N_1,N_2,\cdots\rangle$ (with energy $E=\sum_{n=1} n~N_n$)
except that each $N$ can only be either 0 or 1. The generating function in this
case being,
\beq
Z_F={\rm tr} \e^{\beta E}=\sum_{E=1}d_E~ z^E~,
\eeq
for which there is also an exact expression,
\beq
Z_F=\prod_{n=1} \left(z^0+z^n\right)=
{\sqrt{2}\over z^{1\over 24}}\sqrt{\theta_2(\tau)\over \eta(\tau)}~,
\eeq 
where $\theta_i$, $i=1\cdots 4$ are the Jacobi theta functions.
Considering the modular property 
($\theta_2(-1/\tau)=\sqrt{-i\tau}\theta_4(\tau)$) we find:
\beq
Z_F(z)={\sqrt{2}\over z^{1\over 24}}q^{-{1\over 
24}}\prod_{r={1\over2}}(1-q^{2r})~, ~~~~~q=\e^{2\pi^2\over \log z}~.
\eeq
Using the same method as in the bosonic case (focusing in the $z\to 1$, 
$q\to 0$ limit) we obtain an asymptotic expression suitable to obtaining the 
large $E$ behavior,
\beq
Z_F(z\sim 1)\sim  \sqrt{2}\e^{-{\pi^2\over 12 \log z}}~.
\eeq
Thus,
\beq
d_E\sim{1\over 2\pi i}\oint {\sqrt{2}\over 
z^{E+1}}\e^{-{\pi^2\over 12 \log z}} ~dz~,
\eeq
The stationary point at $\log z=-\pi/2{\sqrt{3(E+1)}}$ yields in this case,
\beq
d_E \equiv {dN \over dE} \sim{1\over 2 ~(3\Delta k)^{1\over4}E^{3\over4}}\e ^{\sqrt{E\over 3\Delta k}
\pi}~,~~~~E\gg \Delta k~,
\label{dEFermi}
\eeq
where we have reinserted the eigenvalue spacing $\Delta k$ in the final step as
 in the bosonic case.

As mentioned in the previous subsection, a compactified spacial direction 
implies the right and left moving modes are independent and, therefore, the 
generating function becomes the product of the two corresponding
factors.

We note that the exponent in the Fermion density of states
(Eqn. \ref{dEFermi}) is a factor of $\sqrt{2}$ smaller than for the
Bose case (Eqn. \ref{dEBose}).  For the huge exponents that concern us
in this article this makes the Fermion density of states highly
subdominant vs the Bose case at the same energy.

\subsection{Bosonization}

Notice that both for fermions and bosons the density of states grows 
exponentially with the square root of the energy. This leads us to the 
following question: is the dominant contribution to the density of states 
coming from states with many modes singly excited? (given that it is the 
only possibility for fermions).     

The similarity in the behavior of the density of states for 
bosons and fermions can be traced to the close relation between bosons and 
fermions in the particular case of $1+1$ dimensions that leads to the concept 
of bosonization.  

The rule that relates a the left-moving part of a boson field $\phi(z)$ and 
the left-moving component of a fermion $\psi(z)$ is
\beq\label{bo}
\psi=\e^{i\phi}~,
\eeq
where we have switched to complex-plane variables $(\sigma, ~t)\to (z,~\bar z)$
defining $z=\rho \e^{i\sigma}$ ($i\log \rho=t$) for later convenience.
Now we take into account the mode expansions,
\beq
\phi (z)=\sum a_n z^{-n}~,~~~~~~ \psi (z)=\sum b_r z^{-r-{1\over 2}}~,
\eeq 
to express the creation operators in the following way
\beq
a_n&=&{1\over 2\pi i}\oint {\phi\over z^{-n+1}}~,\nonumber\\ 
b_r &=&{1\over 2 \pi i}\oint{\psi\over z^{-r+{1\over 2}}}={1\over 2 \pi i}
\oint{\e^{i \phi}\over z^{-r+{1\over 2}}}~.
\eeq
And through Eqn.~\ref{bo} we obtain,
\beq\label{rel}
b_r={1\over 2 \pi i}\oint
{\e^{i \phi}\over z^{-r+{1\over 2}}}={1\over 2 \pi i}\sum_m\oint
{\left(i\sum_pa_p ~z^p\right)^m\over m!~ z^{-r+{1\over 2}}}~.
\eeq
The implications of Eqn.~\ref{rel} can be illustrated by the following 
example:
take a singly excited fermionic state of energy $E=10$, 
\beq\label{example}
b_{-10-{1\over 2}}|0\rangle&=&{1\over 10!}\oint {(a_{-1}z)^{10}\over z^{11}}+
{1\over 5!}\oint {(a_{-2}z^2)^{5}\over z^{11}}\nonumber\\&&
\hskip-1cm+{1\over 2!}\oint 
{(a_{-5}z^5)^{2}\over z^{11}}+\oint {(a_{-10}z^{10})\over z^{11}}|0\rangle
\nonumber\\ 
&&\hskip-1.5cm =\left({1\over 10!}(a_{-1})^{10}+{1\over 5!}(a_{-2})^{5}+
{1\over 2}(a_{-5})^{2}+a_{-10}\right)|0\rangle\nonumber\\
&&\hskip-1.5cm ={1\over\sqrt{10!}}|n_1=10\rangle
+{1\over\sqrt{5!}}|n_2=5\rangle+|n_{10}=1\rangle
\eeq
it is equivalent to a linear combination of multiply excited bosonic states 
of level given by the divisors of $E$. In the last line of 
Eqn.~\ref{example} we 
used normalized states $|n_i=m\rangle\equiv{1\over \sqrt{m!}}
\left(a^\dagger_i\right)^m|0\rangle$.

In general, a highly energetic singly excited fermionic state $b_r|0\rangle$ is
equivalent to a bosonic state of singly excited bosons contaminated by small
components of high multiplicity.

\bibliographystyle{apsrev}
\bibliography{Clock}

\end{document}